\documentclass[10pt,fleqn,twocolumn,oneside]{article}

\usepackage{graphicx}                                                   % package for handling graphics
\usepackage[top=1.0in,bottom=1.0in,left=1.0in,right=1.0in]{geometry}
\usepackage[super,sort&compress,comma]{natbib}

\begin{document}

\title{A charged particle Talbot-Lau interferometer and magnetic field sensing} %Title of paper

\author{Roger Bach, Glen Gronniger, and Herman Batelaan}

\maketitle

\begin{center}\begin{minipage}{0.90\linewidth}
We present the first demonstration of a three grating Talbot-Lau interferometer for electrons.
As a proof of principle, the interferometer is used to measure magnetic fields.
The device is similar to the classical Moir\'{e} deflectometer.
The possibility to extend this work to build a scaled-up charged particle deflectometer or
interferometer for sensitive magnetic field sensing is indicated.
\end{minipage}\end{center}

In the past decades matter-wave interferometers have been used to demonstrate fundamental quantum phenomena
and perform precision measurements.

Applications include accelerometers\cite{Pritchard97,Zeilinger96},
gravity gradiometers\cite{Kasevich06},
detection of decoherence\cite{Arndt04,Hasselbach07},
and the measurement of fundamental constants\cite{Kasevich06,Pritchard02}.
Such work has been mostly carried out with atomic or molecular beam interferometers.
Electron beam interferometers have made use of the presence of charge to demonstrate the Aharonov-Bohm effect\cite{Tonomura98},
to visualize super-conducting vortices, and observe degeneracy in free space\cite{Hasselbach02}.
It is also well-known that electron interferometers share with their electron microscope counterparts
the requirement that external electromagnetic fields need to be  carefully shielded.
In Tonomura's work, nearby commuter trains caused instability, while in the work at Tubingen a radio
station reduced interference contrast.
In view of this, it appears natural to investigate using this sensitivity to our advantage. 
In this paper, we start an investigation into the use of electron interferometry as a magnetic field sensor. 

Different interferometer designs exist. The present discussion is limited to free electron beam interferometers
and does not include the exciting field of mesoscopic or solid state electron interferometry. 
Biprism interferometers\cite{Mollenstedt80,Hasselbach93}, and three grating far-field interferometers have been demonstrated\cite{Gronniger06}.
Larger interferometers are usually more sensitive to fields, but also have to meet stringent mechanical demands.
Here, we opted for a small near-field interferometer.
The recent observation of two-grating Talbot-Lau interference fringe patterns\cite{Cronin06,Cronin09b} motivated the construction of a three grating Talbot-Lau interferometer.
The promise that such an approach offers is mechanical stability in a small design, with a large electron beam acceptance angle. 

In this paper, we demonstrate the first charged particle Talbot-Lau interferometer (TLI).
A TLI is setup with three gratings. The first pair of gratings produces fringes downstream utilizing the Lau effect.
The third grating is added and used as a mask to sample these fringes.

The interferometer is placed in an adjustable magnetic field to scan the fringes. A sensitivity to DC magnetic fields of 9.5 $\mathrm{nT \, Hz^{-1/2}}$
was measured. This is a modest result as compared to what can be reached with conventional devices, e.g., fluxgate magnetometers\cite{Ripka03},
atomic sensors\cite{Edelstein06}, or squids\cite{Clark96},
but is a proof of principle for the operation of the device. The scalability of the TLI magnetic sensor is discussed below.

\begin{figure*}[tb]
\centering
\includegraphics[width=\textwidth]{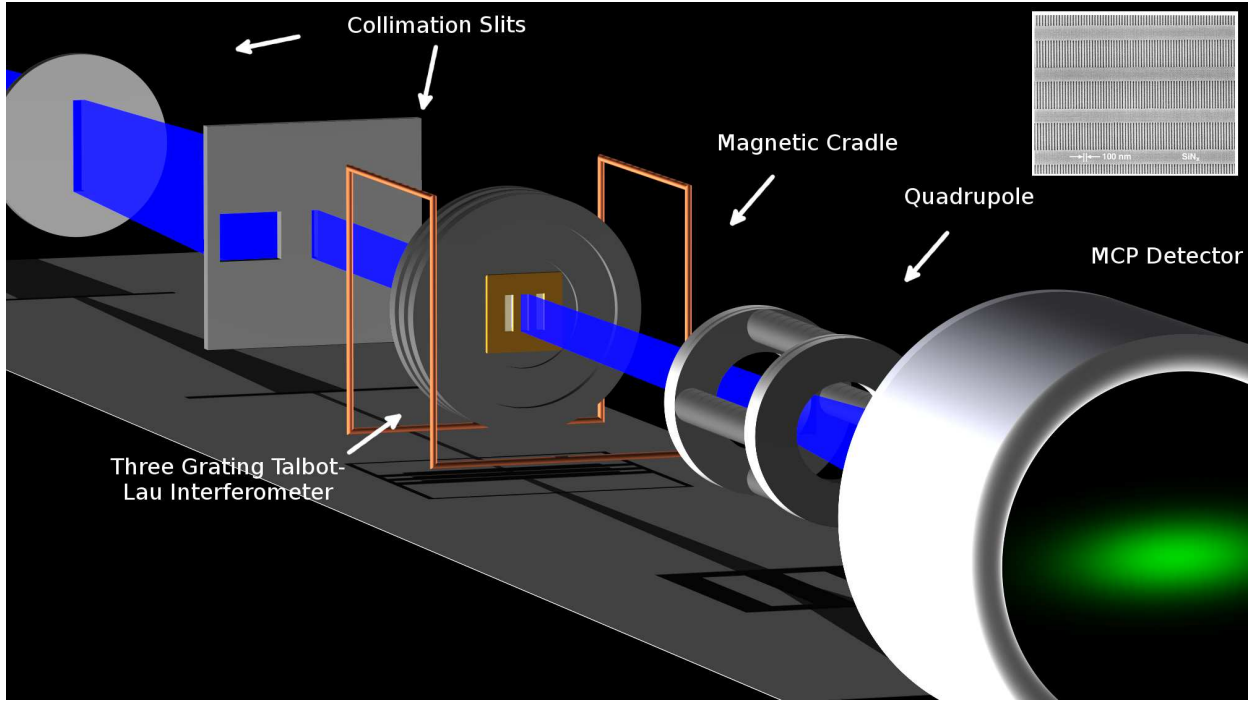}
\caption[Setup]{Illustration of the experimental setup. 
Electrons pass through two collimations slit. The first being 5 $\mu$m wide by 100 $\mu$m tall. 
The second being either a a 2 $\mu$m wide by 10 $\mu$m tall or a 30 $\mu$m wide by 10 $\mu$m tall.
The collimated electron beam then passes through the Talbot-Lau Interferometer (TLI). The TLI consists of three identical section separated by 3.06 $\pm$ .01 mm.
Each section has a metal-coated silicon nitride 100 nm periodic grating. A scanning electron micrograph of the grating structure is shown in the inset.
The total throughput of the TLI is counted by a micro-channel plate (MCP) and phosphorus screen detector. To prevent saturation of individual channels a
electron static quadrupole lens is used to magnify the electron throughput beam spatially. To apply magnetic an electric fields, the TLI was placed in
a cradle with a current carrying wire and between two charged plates (not shown).}
\label{fig:Setup}
\end{figure*}

See Fig. \ref{fig:Setup} for an illustration of the experimental setup. The TLI was placed in a vacuum system held at a pressure below $1 \times 10^{-7}$ Torr
and static external magnetic fields where shielded to better than .5 $\mu$T.
Electrons from a variable energy thermionic emission electron gun (Kimball Physics Egg-3101) are first collimated by a 5 $\mu$m wide by 100 $\mu$m tall slit placed 32 cm from the gun exit. 
For further collimation one of two second collimation slit was chosen.
A 2 $\mu$m wide by 10 $\mu$m tall or a 30 $\mu$m wide by 10 $\mu$m tall slit could be positioned in the beam at a distance of 24 cm from the first slit.
The collimated beam then passed through the three grating Talbot-Lau Interferometer, which is placed at a distance of 5 cm from the second collimation slit.
The total electron transmission through the interferometer is then counted on a two-dimensional micro-channel plate (MCP) and phosphorus screen detector.
The detector is placed 41 cm from the interferometer. An electrostatic quadruple, located 11 cm before the detector, is used to spatially magnify
the electron transmission to prevent saturation of the MCP detector. The quadruple and MCP detector can provide two-dimensional position information,
but in this experiment it was only used for counting and alignment purposes. Individual electron detection events on the MCP detector's phosphorus screen are
then discriminated and sent to counters.

The three grating Talbot-Lau interferometer (see Fig. \ref{fig:Setup}) consists of three identical sections separated by 3.06 $\pm$ .01 mm.
Each section consists of a 30 mm diameter aluminum body with an 8 mm hole in the center. Attached to each section is a metal-coated silicon nitride 100 nm periodic grating.
The grating structures are shown in the inset of Fig. \ref{fig:Setup} and were made by Savas and Smith at the MIT NanoStructures laboratory using
achromatic interferometric lithography.\cite{Savas95,Savas96} The three sections are rotationally aligned to $10^{-3}$ rad by diffracting a HeNe
laser off the 1.5 $\mu$m support structure.

To apply magnetic fields the TLI was placed in a cradle that had a wire arranged around a cube's edges as shown in Fig. \ref{fig:Setup}. This arrangement produces a maximum
magnetic field at the center of the structure in the vertical direction and no field in the horizontal directions. At the center the magnetic field ($B$) is given by
\begin{equation}
B = \frac{4}{\sqrt{3}} \frac{ \mu_0 I }{\pi w};
\label{eq:magnetic}
\end{equation}
where $\mu_0$ is the permeability of free space, $I$ is the current through the wires, and $w$ is the length of the cube edge that the wires are arranged on (54 mm).

\begin{figure}[tbh]
\centering
\includegraphics[width=\columnwidth]{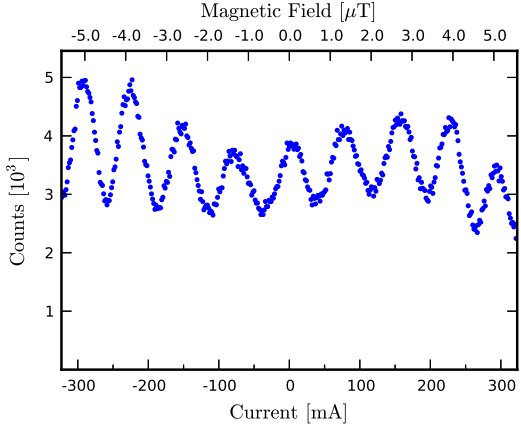}
\caption[Magnetic Dependence]{Electron throughput of the Talbot-Lau interferometer with varying magnetic field.
The current through the magnetic cradle was scanned from a negative to positive value.
Each point represents 10 passes with 100 ms of integration time on each pass. The theoretical maximum magnetic field is shown along the top axis.
The throughput oscillates with a periodicity of 1.2 $\mu$T.}
\label{fig:Magnetic}
\end{figure}

The total electron throughput of the TLI was recorded as a function of current (Fig. \ref{fig:Magnetic}).
The theoretical magnetic field at the center of the cradle, using Eq.~\ref{eq:magnetic}, is shown along the top axis, while the actual
current through the wires is shown along the bottom.
The electron count rate data is the sum of multiple sweeps.
A full period of oscillation of the count rate corresponds to a displacement of one grating period ($d$).
The Fourier transform of the data yields a period of 71 mA or 1.2 $\mu$T.
Using an impulse approximation for the classical deflection of a particle with charge $q$ in a uniform magnetic field is
\begin{equation}
s = \frac{q B L^2}{2 m v} = \frac{q B L^2}{2\sqrt{2 m E}}.
\label{eq:displacement}
\end{equation}
Here $s$ is the transverse displacement, $L$ is the length of the field region, $m$ the mass of the particle, $v$ the velocity of the particle, and  $E$ the energy of the particle.
Alternately, from quantum mechanics, the phase difference can be approximated by\cite{AB59}
\begin{equation}
\phi = \frac{q}{\hbar} \int{ \vec{B} \cdot d\vec{A}} \simeq \frac{q}{\hbar} B L^2 \sin{\theta},
\label{eq:qm}
\end{equation}
where $d \sin{\theta} = \lambda_{dB}$ and $\lambda_{dB}$ is the de Broglie wavelength.
In Fig. \ref{fig:Magnetic} the energy used was 10 keV.
The magnetic field needed to cause a Lorentz force deflection of one grating period is 1.8 $\mu$T.
This 35\% difference is probably due to the TLI not being exactly centered in the cradle and the cradle not being perfectly cubic.

\begin{figure}[hb]
\centering
\includegraphics[width=\columnwidth]{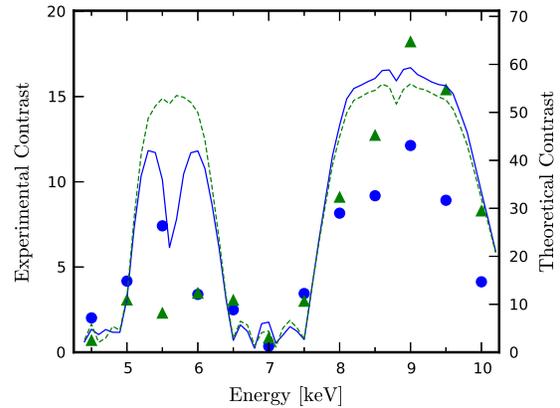}
\caption[Magnetic Dependence]{The experimental (points) and theoretical (lines) contrast of the Talbot-Lau interferometer (TLI) are shown.
The TLI exhibits contrast maxima when the energy of the electrons satisfy the Talbot length (see text). These energies are 8.8 and 5.6 keV. Experimental results using a 2 and 30 $\mu$m second collimation slit are shown using circles (blue) and triangles (green) respectively. 
Similarly theoretical contrasts are shown using solid (blue) and dashed (green) lines. The vertical scales for the experimental and theoretical contrasts
are different and shown on the left and right sides respectively.}
\label{fig:Contrast}
\end{figure}

One key difference between a TLI and a classical Moir\'{e} deflectometer\cite{Zeilinger96} is the energy dependence.
A Moir\'{e} deflectometer has no energy dependence, where as in a TLI there is a strong contrast dependence on energy.
For the current setup the maximum contrast occurs at lengths around multiples of half the Talbot length $L_T = 2 d^2 / \lambda_{dB}$,\cite{Cronin09}
where $d$ is the period of the grating and $\lambda_{dB}$ is the de Broglie wavelength.
In the current experiment the length was fixed at 3.06 mm and the electron gun could reach energies from 4.5 to 10 KeV.
This allowed us to probe two different contrast maximums, at 8.8 and 5.6 keV or de Broglie wavelengths of 13.1 pm and 16.3 pm respectively.
The contrast ($(S_{max}-S_{min})/(S_{max}+S_{min})$) is plotted as a function of energy in Fig. \ref{fig:Contrast}. The maximums are clearly seen in the experimental data.

Another feature of the TLI is that the contrast should be unaffected by the beam width. The TLI can tolerate a large momentum spread in the beam\cite{Cronin09}.
These features can be seen when comparing data with a 2 $\mu$m collimation slit versus the 30 $\mu$m collimation slit, see Fig \ref{fig:Contrast}.
The contrast is mostly unaffected. In theory the TLI could be used without collimation slits.
This is not possible in our setup because the non-perfect rotational alignment of the three gratings would cause different vertical sections of the beam to be out of phase 
and diminish the contrast.

%Theoretical Section

To determine if the contrast of the experimental results was consistent with quantum mechanics and rule out a classical explanation, a quantum mechanical numerical simulation was performed.
The theoretical description of the physical system is based on Feynman's path integral formulation\cite{Feynman48,Barwick06,Caprez09,Bach13}.
The wave function $\Psi(x)$ was propagated from one plane to the next, i.e., the planes of the collimation slits and gratings. This was done by
\begin{equation}
\Psi_{f}(x) = \int K_{i \rightarrow f}(x',x) \Psi_{i}(x') dx'.
\label{eq:Propogation}
\end{equation}
The coordinate system is chosen so that the $x$-axis is horizontal in Fig.~\ref{fig:Setup} and is perpendicular to the beam
propagation axis, which is the $z$-axis. The kernel in Eq.~\ref{eq:Propogation} is given by
\begin{equation}
K_{i \rightarrow f}(x',x) = \exp \left( i \frac{S(x',x)}{\hbar} \right),
\label{eq:Kernel}
\end{equation}
where $S$ is the classical action. The wavefunction propagates in free space between the planes, for which the action is
\begin{equation}
S(x',x) = 2 \pi \frac{\sqrt{ (x'-x)^2 + (z'-z)^2 }}{\lambda_{dB}}.
\label{eq:S_Kernel}
\end{equation}

At the planes, the wavefunction is modified to
\begin{equation}
\Psi_{out}(x) = A(x) \exp(i \phi(x) ) \Psi_{in}(x),
\label{eq:Plane}
\end{equation}
where the amplitude modulation is given by $A$ and $\phi$ is the phase modulation.
For example, at a grating plane
\begin{equation}
A(x) = \sum_{n=-\infty}^{\infty}{ H\left( x - nd + \frac{fd}{2} \right) \times H\left( -x + nd + \frac{fd}{2} \right) },
\label{eq:grating}
\end{equation}
where $d$ is the grating period, $f$ is the open fraction of the gratings, $H$ is the Heaviside function, and $n$ indicates the $\mathrm{n^{th}}$ slit of the grating.
For a full description of the electromagnetic interaction see Barwick \emph{et al.}.\cite{Barwick06} At each plane an image charge potential was added.
At the first two gratings an additional random potential was added. The values describing the interaction were taken from Barwick \emph{et al.}

To incorporate the electron gun into the simulation incoherent sources were added. A point source on the first collimation slit was
propagated through the setup and then incoherently added up with other point sources from the first collimation slit.
The probability distribution was integrated to calculate the throughput of the TLI.

To simplify the simulation the magnetic field was left out, instead the third grating was translated along the x-axis to mimic the deflection from the magnetic field. 
The simulation was performed multiple times with the third grating in different positions. The contrast was determined from the throughput.
This procedure was repeated for different energies.

The theoretical contrast, as a function energy, is shown in Fig. \ref{fig:Contrast} by lines. Two simulations are shown, the solid line (blue) is with a
2 $\mu$m second collimation slit and the dashed line (green) is with a 30 $\mu$m second collimation slit.
The qualitative shape matches the experimental data (points), but the scale of the contrast is different. The simulations represents
the best case scenario of a perfectly shielded environment and ideal rotational alignment.
Estimating a beam height of 33 $\mu$m at the TLI and a misalignment of $10^{-3}$ rad, the contrast would be reduced by a factor of 2.4.
We believe this to be the dominant contribution to the discrepancy between the experiment and theory.
The only fit parameter was the open fraction.
An open fraction of 35\% best represented the two contrast maximums, where as the gratings were originally manufactured to be 50-60\%.

%Section Describing small field measurements

\begin{figure}[tbh]
\centering
\includegraphics[width=\columnwidth]{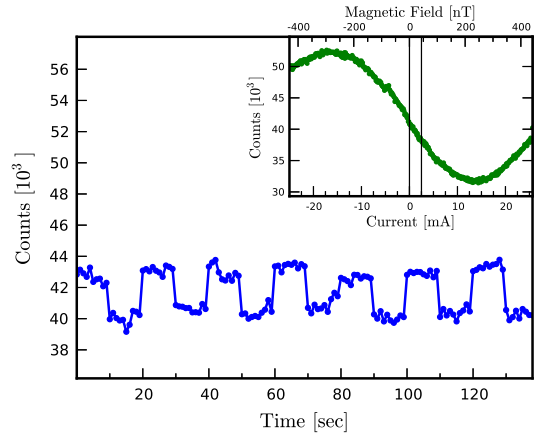}
\caption[Small Field]{To demonstrate the measurement of a magnetic field, a small current was turned on then off repeatedly. The main graph shows
the throughput of the Talbot-Lau interferometer (TLI). A current of 2.5 mA, which corresponds to a field of 43 nT, was tuned on for ten seconds then left off for 10 seconds.
Each point represents 1 sec of integration. The inset shows a larger scan of the current. An electric field was used to place the TLI's throughput at the largest slope.
The two vertical lines represent the currents used during the magnetic field measurement}
\label{fig:Step}
\end{figure}

A small magnetic field was measured to demonstrate the sensitivity of the TLI.
A set of parallel plated placed the TLI in an electric field of approximately 200 V/m to shift the
throughput to halfway between the maximum and minimum. At this position the electron throughput was the most sensitive to a change in magnetic field,
as seen in Fig.~\ref{fig:Step} inset. In the inset the magnetic field is scanned to show almost a full period. Each point is 12 passes with
100 ms of integration time, and then is normalized to represent one second of count rate.

A 2.5 mA current was turned on for 10 seconds and then left off for 10 seconds. The electron throughput is shown in Fig.~\ref{fig:Step}.
Each point represents one second of integration. The current corresponds to a field of 43 nT.
The signal to statistical noise in Fig.~\ref{fig:Step} is approximately 4.5 for a measurement of 1 second duration.
The sensitivity is thus about 9.5 $\mathrm{nT \, Hz^{-1/2}}$. The sensitivity would be limited by statistical noise at 4.7 $\mathrm{nT \, Hz^{-1/2}}$. 

Several steps can be taken to scale the device. The electron beams become separated at a grating distance of 20 mm.\cite{Gronniger06}
For a separation of 10 mm the device length $L$ increases by $10/3$. For an improved rotational alignment, the full surface
of the grating $A$ (1 mm $\times$ 3 mm) could be used at the same electron beam density to increase the throughput by (3 mm / 10 $\mu$m)$\times$(1 mm / 30 $\mu$m).
Finally a typical weak iron magnetic field concentrator can improve the magnetic flux by a factor $C$ of 20.\cite{Kitching94}
As the sensitivity scales with $L^2 C$ (see eq~\ref{eq:displacement} and \ref{eq:qm}) and the throughput should increase the sensitivity by $A^{1/2}$.
An expected performance of a scaled device would be 430 $\mathrm{fT \, Hz^{-1/2}}$.
Additionally, the device works in bias fields exceeding the Earth magnetic field, while its frequency response remains to be explored.

In summary, a Talbot-Lau electron interferometer has been demonstrated. The devise acts as a magnetometer with a modest sensitivity of 9.5 $\mathrm{nT \, Hz^{-1/2}}$,
but appears to be scalable to much better values. The use of charged particle interferometry as an alternative means to magnetic field sensing
as a proof of principle is clear and appears interesting, because its parameters space remains largely unexplored.

%\begin{acknowledgments}
% Put your acknowledgments here.
%\end{acknowledgments}

\bibliographystyle{aipnum4-1}
\bibliography{TL_bib,DS_bib}

\end{document}